\documentclass[a4paper,11pt]{article}
\pdfoutput=1
\usepackage{jcappub}
\usepackage[T1]{fontenc}
\usepackage{enumerate}
\usepackage[mathscr]{euscript}
\usepackage[toc, page]{appendix}
\usepackage{hyperref}
	\hypersetup{
    colorlinks=true,
    linkcolor=blue,
    filecolor=magenta,      
    urlcolor=cyan,
    }
\usepackage{array}
\usepackage{flafter}
\usepackage{layout}
\usepackage{epstopdf}
\usepackage[ansinew]{inputenc}
\usepackage{xfrac}
\usepackage{algorithm}
\usepackage[dvipsnames]{xcolor}

\makeatletter
\gdef\@fpheader{}
\makeatother

\title{Is Cosmological Tuning Fine or Coarse?}

\author[a,1]{Daniel Andr\'es D\'{\i}az-Pach\'on\note{Corresponding author}} 
\author[b]{Ola H\"ossjer} 
\author[c]{Robert J. Marks II}

\affiliation[a]{University of Miami, Division of Biostatistics, Don Soffer Clinical Research Center, 1120 NW 14th St Room 1057, Miami, FL 33136, U.S.A.}
\affiliation[b]{Stockholm University, Dept. of Mathematics, Division of Mathematical Statistics, Roslagsv\"agen 101, Kr\"aftriket, hus 6 Room 318, Stockholm, Sweden}
\affiliation[c]{Baylor University, Dept. of Electrical \& Computer Engineering, One Bear Place \#97356, Waco, TX 76798-7356, U.S.A.}

\emailAdd{Ddiaz3@miami.edu}
\emailAdd{ola@math.su.se}
\emailAdd{Robert\_Marks@Baylor.edu}

\abstract{The fine-tuning of the universe for life, the idea that the constants of nature (or ratios between them) must belong to very small intervals in order for life to exist, has been debated by scientists for several decades. Several criticisms have emerged concerning probabilistic measurement of life-permitting intervals. Herein, a Bayesian statistical approach is used to assign an upper bound for the probability of tuning, which is invariant with respect to change of physical units, and under certain assumptions it is small whenever the life-permitting interval is small on a relative scale. The computation of the upper bound of the tuning probability is achieved by first assuming that the prior is chosen by the principle of maximum entropy (MaxEnt). The unknown parameters of this MaxEnt distribution are then handled in such a way that the weak anthropic principle is not violated. The MaxEnt assumption is ``maximally  noncommittal with regard to missing information.''  This approach is sufficiently general to be applied to constants of current cosmological models, or to other constants possibly under different models. Application of the MaxEnt model reveals, for example, that the ratio of the universal gravitational constant to the square of the Hubble constant is finely tuned in some cases, whereas the amplitude of primordial fluctuations is not.}


\begin{document}
\maketitle
\flushbottom

\section{Introduction}

The fine-tuning of the universe for life is a claim that the constants in the laws of nature or the ratios thereof, and/or the boundary conditions, both in particle physics and in the standard cosmological model, belong to minuscule life-permitting intervals such that outside them life could not exist. The modern version of fine-tuning was introduced by Carter \cite{Carter1974}. Since inception, fine-tuning remains a hot topic among scientists and popularizers \cite{Davies1982, Hoyle1982, Hawking1988, Weinberg1994}.  Perhaps fine-tuning's biggest claim to fame came with the book {\sl The Anthropic Cosmological Principle}, by Barrow and Tipler \cite{BarrowTipler1988}. Subsequently the argument has been scrutinized in physics and cosmology. See, for instance, \cite{TegmarkEtAl2006, LewisBarnes2016, Adams2019} and all the references therein.

Even though the definition of fine-tuning appears to be simple, there are variations in the particulars. On one hand, researchers have differing opinions inasmuch as the constants of nature that must be considered. Just to mention one instance out of many, Adams \cite{Adams2019} and Rees \cite{Rees2000} consider the gravitational constant, but Tegmark et al \cite{TegmarkEtAl2006} and Barnes \cite{Barnes2020} ignore it, limiting themselves to the constants in the Standard Models. On the other hand, when talking about life, consensus is illusive largely because there is not even a consistent definition of what life is. As an illustration, Adams extends his argument to any conjectured form of life, not necessarily carbon-based \cite{Adams2019}; Sandora focuses on complex intelligent life \cite{Sandora2019a, Sandora2019b, Sandora2019c, Sandora2019d}; and philosopher Robin Collins talks about embodied moral agents \cite{Collins2012}.

However, in spite of fine-tuning ado and the passions elicited in critics and defenders, the degree of tuning, either fine or coarse, remains unsettled. To emphasize, the claim is not that the tuning is fine ---or coarse. The claim is that, up to this point, the degree of tuning has yet to be determined. In order to explain this assertion, let's examine in more detail how the tuning should be measured. Assuming the constants to be considered are agreed upon, the process can be summarized in two steps.
\begin{enumerate}
\item The bounds for the life-permitting intervals of constants or of their ratios must be identified.
\item Probabilities over life-permitting intervals must be calculated.
\end{enumerate}
Both steps require elaboration. The main thing to notice is that the first step belongs exclusively to the realm of physics, whereas the second step belongs to the realm of mathematics.

For step 1 physicists have proposed useful boundaries of the life-permitting intervals for many of the constants or the ratios of constants \cite{LewisBarnes2016}. We must add, however, that in the absence of a theory of everything the task is far from complete. Many of these numbers could change in future research in at least three ways: (i) by changing the limits of the life-permitting intervals, (ii) by removing constants because they have lost relevance, or (iii) by identification of constants hitherto not considered.

We focus on step 2 ---finding the degree of tuning. More specifically, in order to determine the probability of having a constant of nature or the ratio between two constants of nature within a life-permitting interval, a Bayesian approach is used. Either the prior distribution of the constant is chosen by the principle of maximum entropy (MaxEnt), or two constants forming a ratio are chosen to have two separate MaxEnt distributions. Such an approach allows finding an upper bound, $P_{\mbox{\scriptsize max}}$, for the probability of tuning. This upper bound is dimensionless, i.e.\ invariant with respect to changing physical units. It is very small for some (but not all) examples, and we refer to tuning as being fine or coarse depending on whether $P_{\mbox{\scriptsize max}}$ is small or not. Our proposed approach is sufficiently general to be applied to the constants in the current (standard) models or to different constants if they change by any of the three aforementioned circumstances.

\section{Probabilities and Tuning}

We do not consider the first step, finding the bounds of life-permitting intervals, relying instead on what has already been identified. Our analysis is restricted to the second step ---how to calculate the probabilities of tuning given that the life-permitting intervals have been determined. It appears that this problem is still open. For instance, Adams halts his presentation at the point of finding the life-permitting intervals, explaining that not much can be said of the probability distributions that must govern the behavior of the constants of nature \cite[p. 6]{Adams2019}:
            \begin{quotation}
            	\noindent A full assessment of fine-tuning requires knowledge of these fundamental probability distributions, one for each [constant of nature] of interest (although they are not necessarily independent).  These probability distributions, however, are not 			currently assessable.
            \end{quotation}
In the absence of these probabilities, not much can be said about the degree of tuning. As surprising as a small interval might intuitively seem, the interval size matters less than the probability over the interval. For instance, for a standard normal distribution, the intervals $(-\infty, 0)$ and $(-0.6745, 0.6745)$ each have probability 0.5 even though the first has infinite length and the second has length 1.349. In contrast, under this distribution, the interval $\left(-\infty, -10^9\right)$, has infinite length but almost zero probability.

In addition, there is the normalization problem. Normalization imposes limitations to Bernoulli's {\em Principle of Insufficient Reason} (PrOIR), otherwise known as the {\em Principle of Indifference} \cite{Bernoulli1713, DembskiMarks2009a, Tschirk2016}. As originally conceived, the PrOIR states that in the absence of any prior knowledge events must be uniformly distributed. The unspoken assumption is that the space must be finite. Normalization posits a general warning against using the PrOIR beyond finite spaces. Placing a uniform distribution over the whole real line, for example, is untenable. McGrew, McGrew, and Vestrup \cite{McGrewMcGrewVestrup2001}; as well as Colyvan, Garfield, and Priest \cite{ColyvanGarfieldPriest2005}; raise the normalization objection in the context of fine-tuning. Since the space where the constants of nature or their ratios could take values has infinite length, attempts to find the probabilities using a uniform distribution cannot be successful. This criticism against the PrOIR in the fine-tuning literature is legitimate.

Yet some persist, for instance, borrowing an idea from quantum field theory in his analysis of the initial entropy of the universe, Roger Penrose reduces an infinite-dimensional phase space $\mathcal P_\mathcal U$ to a new finite-dimensional space in which each dimension has finite size.\footnote{With respect to converting the space to a finite-dimensional one Penrose writes ``In fact $\mathcal P_\mathcal U$ will be infinite dimensional... This causes some technical problems for the definition of entropy, since each required phase space region $\mathcal V$ will have infinite volume. It is usual to deal with this problem by borrowing ideas from quantum (field) theory, which enables a finite answer to be obtained for the phase-space volumes which refer to systems that are appropriately bounded in energy and spatial dimension... Although there is no fully satisfactory way of dealing with these issues in the case of gravity---owing to a lack of a satisfactory theory of quantum gravity---I am going to regard these as technicalities that do not affect the general discussion of the issues raised by the second law.'' \cite[pp.\ 700-701]{Penrose2004}. See also the more formal result in \cite{Penrose1981}. And regarding making each dimension to have finite size he adds: ``We shall use the phase space $\mathcal P_\mathcal U$ appropriate to the entire universe, so the evolution of the universe as a whole is described by the point $x$ moving along a curve $\xi$ in $\mathcal P_\mathcal U$. The curve $\xi$ is parametrized by the time coordinate $t$, and we can expect that, from the second law, $\xi$ enters immensely larger and larger coarse-graining boxes as $t$ increases. We suppose that some `reasonable' coarse graining has been applied to $\mathcal P_\mathcal U$, but if we wish to obtain finite values for the entropies that $x$ encounters, we would want the volumes of these boxes to be finite.'' \cite[pp.\ 701-702]{Penrose2004}.}

Tegmark et al have also addressed the second step of tuning \cite{TegmarkEtAl2006}. They propose to calculate the probability that a constant of nature $x$ belongs to its life-permitting interval by decomposing the density function of $x$ as a product of a prior density and a selection probability  ($f(x) \propto f_{prior}(x)f_{sel}(x)$). Here the prior is a theoretically predicted distribution at some random point during inflation and the selection distribution obtains the probability of observing that point. This second term is related to the weak anthropic principle. Observers in a universe are bound to measure constants of nature for which a habitable universe is possible with a nonzero $f_{sel}$. As an example, Tegmark et al. consider the probability distribution of the vacuum energy density assuming uniformity. However, other distributions are possible \cite{MartelEtAl1998}.

Recently, Barnes has fleshed out these probabilities using a Bayesian approach \cite{Barnes2011, Barnes2017, Barnes2018, LewisBarnes2016}. Barnes, as Collins \cite{Collins2012}, does what McGrew, McGrew, and Vestrup criticize in \cite{McGrewMcGrewVestrup2001} by assuming finite sample spaces, therefore using uniform distributions to calculate probabilities. To our knowledge, Barnes has been the first to justify his use of a continuous uniform (or what physicists call a ``flat'' distribution which is the MaxEnt distribution in a space of finite size absent all other knowledge), following Edwin Jaynes's recommendations \cite{Jaynes2003}.

Throughout this article we also apply a Bayesian approach \cite{Berger1985}.  Using informational maximum entropy, a concept also due to Jaynes \cite{Jaynes1957a, Jaynes1957b},  we directly assume an infinite sample space for constants of nature, or ratios of such constants, using a class of prior distributions over this space. The maximum entropy principle generalizes the PrOIR. Maximum entropy over an infinite domain is no longer uniform but is still applicable under the appropriate MaxEnt prior distribution, given some restrictions.  

A clarification should be made that entropy as treated here refers to information entropy, not to thermodynamics entropy. In thermodynamics, one uses entropy in a slightly different way than here: Randomness refers to the {\it degree of disorder of a physical system}, and the second law of thermodynamics states that the entropy of a closed system, without external influences, will increase towards an equilibrium state of maximal disorder. In this paper randomness refers to {\it epistemic uncertainty} regarding the value of a constant of nature, or the ratio between two such constants. In particular, entropy in information is an inherent property of the distribution function that describes this epistemic uncertainty, and it corresponds to our degree of ignorance. This perspective allows to extend entropy even to distributions of random variables whose domain is not of finite (counting or Lebesgue) measure.

Then, with appropriate restrictions applied to the moments of distributions in non-compact spaces, it is possible to find even in these settings MaxEnt distributions. In particular, for our case of interest, maximum entropy for the distribution of the constants of nature or ratios then refers to a maximum degree of ignorance about their values, given the imposed restrictions. As Jaynes saw it, entropy in statistical mechanics is but an application of the more general information concept \cite{Jaynes1957a}. 

Even though our main focus in this article is epistemic uncertainty, it should be added that MaxEnt distributions for physical systems are observed in nature. As we saw, the second  law  of  thermodynamics  famously states that the entropy of a gas in a closed room reaches maximum entropy in pressure in accordance to a uniform PrOIR. What if, on the other hand, there is only a single boundary? The barometric pressure measured from the surface of the earth to space follows the MaxEnt distribution of  Exp(1/$\mu$).  An example of MaxEnt that is unbounded for negative and positive values is the Maxwell-Boltzmann distribution describing the velocity of ideal gas particles. The projection of the velocity vector along each direction is unbounded with a $\mathcal N(0,\sigma^2)$ distribution, which is  MaxEnt for domains bounded on neither side (see Table 1 below).

\section{Tuning and MaxEnt}

Fine-tuning asserts that the constants of nature and the boundary conditions of the universe must live in narrow intervals of low probability in order to make life possible and that if such constants would not have had the actual values they possess, life would have never been possible. Therefore, fine-tuning arguments assume that the value of a constant of nature, or the ratio between two constants, let's call it $x=X_\text{obs}$, is an observation of a random variable $X$. Then, for $X$, a probability of the life-permitting interval is calculated.

As seen in the previous section, the interval width of the life-permitting interval is less important than the corresponding probability. Denote the life-permitting interval of $X$ as $\ell_X$, and its length as $|\ell_X|$. In order to properly assess this probability, three steps (I-III) need to be followed:

\vspace{0.1 in}
\textbf{I) Determine the {\it right} sample space $\Omega$ for $X$}. The incorrect determination of $\Omega$ is what McGrew, McGrew, and Vestrup criticized, and ad-hoc attempts to force a finite sample space from an infinite one do not seem convincing \cite{McGrewMcGrew2005}. Proceeding is not possible without first providing the right sample space. 

\vspace{0.1 in}
\textbf{II) Find a probability distribution $F$ of $X$}. $F$ must be such that it best represents the current knowledge of the behavior of a constant of nature or of the ratio between two constants in the most noncommittal way. In this step we apply, instead of the uniform assumption of the PrOIR, either the MaxEnt principle to $X$ itself or the MaxEnt principle applied to each of the two constants whose ratio is $X$. Jaynes  summarizes the advantages of MaxEnt over the PrOIR in this way \cite[p. 623]{Jaynes1957a}:

\begin{quotation}
	\noindent The principle of [MaxEnt] may be regarded as an extension of the [PrOIR] (to which it reduces in case no information is given except enumeration of possibilities $x_i$), with the following essential difference. The [MaxEnt] distribution may be asserted for the positive reason that it is uniquely determined as the one which is {\it maximally noncommittal with regard to missing information}, instead of the negative one that there was no reason to think otherwise. Thus the concept of entropy supplies the missing criterion of choice which Laplace needed to remove the apparent arbitrariness of the [PrOIR], and in addition shows precisely how this principle is to be modified in case there are reasons for ``thinking otherwise.'' [Emphasis added.]
\end{quotation}

The MaxEnt principle can be applied directly to $\Omega$ for the case of consideration of a single constant. Another option is to assume that in the ratio of two physical constants, each constant is MaxEnt. MaxEnt can still find a reference distribution for unbounded domains \cite{Conrad2005, CoverThomas2006}, whereas the PrOIR cannot. Table \ref{maxent} below illustrates the best-known cases of MaxEnt distributions that can be applied to constants or ratios of constants living in a unidimensional space. (A more comprehensive table can be found in \cite{ParkBera2009}.)

To introduce the MaxEnt Bayesian approach, first consider a problematic application to fine-tuning. Let $x$ be an observation of a random variable $X$ that belongs to $\Omega$. Since the sample size is 1, one may first assume $x$ was chosen as the observed value of $X$, and, as such, $x = X_\text{obs}$ would become the average of the sample. Being a sample of size one, this value serves as a sufficient statistic for the expected value $\mu=E(X)$ of $X$ \cite[Ch.~6]{CasellaBerger2006}. In usual statistical notation, $\hat \mu = x$. From the first step the space is known and, from the second step, an estimated value for the mean can be obtained.

However, this approach has a weakness due to the weak anthropic principle. Since we live in a habitable universe, $x$ is not an unbiased observation of a random variable $X$ with distribution $F$. It is rather an observation of a random variable with a truncated distribution
\begin{equation*}
	F_\text{trunc} = F | X\in \ell_X.
\end{equation*}
Then $\hat \mu = x$ is not an unbiased estimate of $\mu=E_F(X)$ but rather an unbiased estimate of $\mu^* = E_{F_\text{trunc}}(X)$.  In particular, if $\ell_X$ is narrow, so that a uniform distribution approximates $F_\text{trunc}$, then $\mu^*$ approximately equals the mid point of $\ell_X$. On the other hand, $F_\text{trunc}$ may differ substantially from a uniform distribution for constants of nature with a wider $\ell_X$. In such a case $x$ is not an unbiased estimate of the mid point of $\ell_X$.

These considerations lead to the proper approach: although $x$ is not an observation of $X \sim F$, we can assume that $X$ belongs to a class of MaxEnt distributions $F=F( \; \cdot \ ;  \theta)$ taken from Table \ref{maxent}, with $\theta = (\theta_1,\ldots,\theta_d)$ a finite-dimensional unknown parameter that due to the weak anthropic principle cannot be estimated easily. In Bayesian statistics such a parameter on the prior distribution $F$ is referred to as a hyperparameter \cite{Berger1985}. Alternatively, when $X=G/D$ is the ratio of two constants of nature $G$ and $D$, we can also reasonably assume that $G$ and $D$ are independent with distributions $F_G=F_G(\; \cdot \ ; \theta_G)$ and $F_D=F_D(\; \cdot \ ; \theta_D)$ chosen as MaxEnt distributions from Table \ref{maxent}. We need then to find the distribution $F(\; \cdot \ ; \theta)$ of $X$, where $F$ and the hyperparameter $\theta$ are functions of $F_G$, $\theta_G$, $F_D$, and $\theta_D$. In both cases, whether a MaxEnt distribution is used  for $X$ itself, or for $G$ and $D$, the MaxEnt principle helps to reduce the class of possible $F$ from an infinite class of distributions to a finite-dimensional class of distributions.

	\begin{table}[h!]
		\centering
		\begin{tabular}{| c | c | c|}
			\hline
			Space & Restrictions/Knowledge & MaxEnt distribution\\
			\hline
			Finite & None & Equiprobability \\
			Finite interval $[a,b]$ & None & $\mathcal U(a,b)$ \\
			Finite interval $[a,b]$ &  $\textbf EX=\mu_T$; $\textbf E(X-\mu_T)^2= \sigma_T^2$ & Truncated normal \\
			$\mathbb N$  & $\textbf EX = \mu$ & Geom($1/\mu$)\\
			$\mathbb R^+$ & $\textbf EX = \mu$ & Exp($1/\mu$)\\
			$\mathbb R$ & $\textbf EX=\mu$; $\textbf E(X-\mu)^2= \sigma^2$ & $\mathcal N(\mu,\sigma^2)$\\
			\hline
		\end{tabular}
		\caption{Maximum entropy distributions over some relevant spaces under different restrictions.}\label{maxent}
	\end{table}

\vspace{0.1 in}
\textbf{III) Calculate the maximum probability of the life-permitting interval $\ell_X$ under the class of distributions 
$\mathcal F = \{F(\cdot;\theta); \, \theta\in\Theta)\}$.} Following Thorvaldsen and H{\"o}ssjer \cite{ThorvaldsenHossjer2020}, let $A$ be the event ``We observe a universe that exists and permits life,'' and let us now regard $x$ as a parameter for a model $P(A|x)$ that gives the probability of observing a life-permitting universe. The tuning probability of the event $A$ is then obtained by regarding $x$ as an observation of $X\sim F$, and averaging $P(A|x)$ with respect to $x$, i.e.,
\begin{equation}\label{TwoPrime}
P(A;\theta) = \int_{\Omega} P(A | x)dF(x;\theta ),
\end{equation}
where, provided the prior density $f(x;\theta)$ exists, $dF(x;\theta) = f(x;\theta)dx$; and $P(A | x)$ is the likelihood. If $A=\ell_X$, then $P(A | x) = 1$ if $x \in A$, and $P(A | x) = 0$ if $x\notin A$. Then (\ref{TwoPrime}) reduces to the tuning probability
\begin{equation}\label{ThreePrime}
P(A;\theta) = F(\ell_X ;\theta)
\end{equation}
considered here. Notice in particular that we might know of $\ell_X$ through the experiments or research that led to $\ell_X$ being determined. However, in any case, the event $A$ is well defined whether we, as observers, know of $\ell_X$ or not. 

In order to finalize the third step and determine the degree of tuning, we also need to maximize (\ref{ThreePrime}) with respect to the hyperparameter $\theta$, when $\theta$ varies over a finite-dimensional space $\Theta$. That is, our final degree of tuning equals 
\begin{equation}\label{Pmax}
	P_{\mbox{\scriptsize max}} = \max_{\theta\in\Theta} F(\ell_X;\theta).
\end{equation}

Notice in particular that  the degree of tuning \eqref{Pmax} can be calculated without violating the weak anthropic principle. This is so, since we did not assume that $x$ is an observer's value of $X$ (since this would have enforced $X\sim F_\text{trunc}$), nor did we estimate the hyperparameter $\theta$ from the single observation $x$, but rather maximized the tuning probability \eqref{ThreePrime} with respect to $\theta$. 

Let us from now on assume that $x$ is the mid point of the life permitting interval. In more detail, suppose the life permitting interval 
\begin{equation}\label{lX}
	\ell_X = x [1-\varepsilon,1+\varepsilon] = [a,b]
\end{equation}
is centered around $x$, with a relative half size $\varepsilon$. In the next section, it will be shown that whenever $\ell_X$ is small, the upper bound (3) of the tuning probability is proportional to $\varepsilon$, i.e.
\begin{equation}\label{PmaxEps}
	P_\text{max}(\varepsilon) = C\varepsilon,
\end{equation}
for a constant of proportionality $C$ that depends on the size of the family of prior distributions $\mathcal F =\{F(\cdot;\theta); \, \theta\in\Theta\}$. Since $\varepsilon$ is a dimensionless constant, it follows that the upper bound of the tuning probability is dimensionless as well.  

Now, in practice the life-permitting interval $\ell_X=[a,b]$ will never be centered around the actual value of the physical parameter. It is possible to solve this inconvenience by retaining our assumption $x=(a+b)/2$, but not requiring $x=X_\text{obs}$. However, for simplicity of exposition we will not make any distinction between $x$ and $X_\text{obs}$ in the text. 

In the next two sections the proposed methodology is applied to the gravitational constant and the primordial fluctuations using values of the width of life-permitting intervals found in the literature. In both cases we assume no additional knowledge beyond the sample space $\Omega$ of $X$, and one or two moments of $X$. This assumption will simplify the calculation of the respective probabilities. It does not mean, however, that additional justified restrictions on the distribution of $X$, based on various types of knowledge, cannot be incorporated. In Section 6.3 we argue that in order to determine the parametric class $\mathcal F$ of possible distributions of $X$, it is always possible to apply the maximum entropy principle {\sl conditionally} on the restrictions that were imposed on $X$.

\section{Example 1: The Gravitational Constant}

	The quantity of interest here is the gravitational constant $G_\text{obs} = 6.67408 \times 10^{-11} \text m^3 \text{kg}^{-1} \text s^{-2}$. Thus $G_\text{obs}$ is an observation of the random variable $G$. However, in this example $G_\text{obs}$ itself is not considered, but rather a ratio $x=G_\text{obs}/d$ between $G_\text{obs}$ and some other constant of nature $d = D_\text{obs}$. This ratio is then an observation of the random quantity $X=G/D$.  
	
	Several possible life-permitting intervals of the ratio $x=G_\text{obs}/d$ between $G_\text{obs}$ and some other constant of nature $d$ are now illustrated. In all of these cases the life-permitting interval has the form defined in (\ref{lX}); i.e.,
	\begin{equation}\label{intervalell}
		\ell_X = [x-\delta, x+\delta] = x\cdot [1-\varepsilon,1+\varepsilon],
	\end{equation}
	where $\delta$ is a positive number usually small, and $\varepsilon = \delta/x$, half of the relative size, is a dimensionless small number. As we will see below, our tuning results will confirm (\ref{PmaxEps}) and be expressed in terms of this dimensionless $\varepsilon$.

	When $D=H^2$ is the random variable corresponding to the Hubble's constant squared, with observed value $d = H^2_\text{obs}$, the first Friedmann equation (assuming $\Lambda_\text{bare}$, the cosmological constant, to be 0) is $x = G_\text{obs}/H^2_\text{obs} = 3/(8\pi\rho_\text{crit})$, where $\rho_\text{crit}$ is the critical density of the universe. According to Davies, it then follows  that $\varepsilon = 10^{-60}$ \cite[pp.\ 88-89]{Davies1982}. 

	Let $d = \Lambda_\text{vac}$, the dynamical contribution from vacuum energy to the cosmological constant. Under the Weinberg-Salam electroweak theory, $G_\text{obs}/\Lambda_\text{vac} = -\sqrt{2}c^4g_w/\pi m_\phi^2$, where $c$ is the speed of light, $g_w$ is the weak force constant and $m_\phi$ is the mass of the scalar particles. In this scenario, Davies suggests $\varepsilon = 10^{-50}$ or even $10^{-100}$ in the case of grand unified theories  \cite[p.\ 107]{Davies1982}. In fact, taking $\Lambda_\text{bare} \neq 0$, we arrive at the cosmological constant problem in which the observed total cosmological constant, $\Lambda_\text{tot} = \Lambda_\text{bare} + \Lambda_\text{vac}$ is $10^{120}$ times smaller than the predicted value of $\Lambda_\text{vac}$ \cite[Ch.\ 5]{LewisBarnes2016}.
	
	Other ratios are possible. For instance, the ratio of constant of gravity to the constant of electromagnetic force is common in the fine-tuning literature \cite[Ch.\ 4]{LewisBarnes2016}, \cite[Ch.\ 3]{Rees2000}, \cite[Sect.\ IV]{Uzan2011}. Whatever the ratio, or the constant, the general three-step method can be applied. Once the life-permitting interval $\ell_X$ has been determined, (\ref{Pmax}) can be used to obtain the tuning probability.

	\textbf{Remark:} Notice that, even though we are using $G_\text{obs}$ here, we can equivalently evaluate the tuning of the gravitational fine-structure constant $\alpha_G = G_\text{obs}m_P^2/(\hbar c) \approx 5.9 \times 10^{-39}$, where $m_P$ is the Planck mass, $\hbar$ is the reduced Planck constant, and $c$ is the speed of light. In contrast to $G_\text{obs}$, $\alpha_G$ has the advantage of being dimensionless. Whereas $G_{\mbox{\scriptsize obs}}$ has only been accurately measured to a relative precision of $\varepsilon=10^{-15}$ \cite{XueEtAl2020}, the fine-tuning assertions of $\alpha_G$ correspond to relative precision of $\varepsilon=10^{-60}$. In any case, $\varepsilon$ is dimensionless for both of $G$ and $\alpha_G$. 
	
	In the next subsections we study the tuning status of the gravitational constant. We do so under different assumptions of the relevant space, which in turn reverberates in the election of the underlying MaxEnt distribution.

\subsection{The maximum entropy principle applied to a ratio $G/D$ in $\mathbb R^+$.}

	According to the first step, in order to determine the appropriate sample space, gravity is assumed to be an attraction force, i.e., $G_\text{obs} > 0$, and the other constant of nature $d$ is non-negative. Therefore $x$ must be a non-negative real number as well. Thus the sample space to be considered is $\Omega = \mathbb R^+$.

	For the second step ---determining the right distribution---  recall that the untruncated distribution $F$ of $X$ is of interest. Since $X$ takes on values in $\mathbb R^+$, 
	\begin{equation}\label{GravDistrib}
		P(X\le z) = F(z;\mu) = F\left( \frac{z}{\mu} ;1 \right)
	\end{equation}
	for some distribution $F$ (chosen below) with scale parameter $\theta=\mu$. Thus, we choose the distribution of $X$ according to the MaxEnt principle. Since the sample space is $\Omega= \mathbb R^+$, it follows from Table \ref{maxent} that $F\sim \mbox{Exp}(1/\mu)$. 
	
	As for the third step, the probability of tuning is:	
	\begin{align}\label{Two}
		P(X\in  \ell_X) &= \exp(-(x-\delta)/\mu) - \exp(-(x+\delta)/\mu) \nonumber \\
				&= \exp(-(x-\delta)/\mu)[1-\exp(-2\delta/\mu)] \nonumber \\
				&= 2e^{-x/\mu} \sinh (\delta/\mu) \nonumber \\
				&= P(\mu,\varepsilon),
	\end{align}
	where $\varepsilon = \delta/x$. Although $P(\mu,\varepsilon)$ depends on $\mu$, it is uniformly small in $\mu$ (while keeping $\varepsilon$ fixed). That is, according to \eqref{Pmax}, and in agreement with \eqref{PmaxEps}, the degree of tuning,
	\begin{equation}\label{Three}
		P_\text{max}(\varepsilon) = \sup_{\mu > 0} P(\mu,\varepsilon),
	\end{equation}
	is of the same order as $\varepsilon$.  $P_\text{max}(\varepsilon)$ in (\ref{Three}) can be calculated analytically by first solving for $dP(\mu,\varepsilon)/d \mu=0$, and then inserting this value of $\mu$ into (\ref{Two}). In fact, $P(\cdot;\varepsilon)$ has a 		maximum at
	\begin{equation*}
		\mu = \frac{2x\varepsilon}{\log((1+ \varepsilon)/(1 - \varepsilon))}.
	\end{equation*}
	A Taylor expansion of the denominator of the maximizing $\mu$ gives
	\begin{equation}
		\mu \approx x(1-\varepsilon) \approx x,
	\end{equation}
	so that
	\begin{align}\label{ExMax1}
		P_\text{max}(\varepsilon) &\approx P(x;\varepsilon) \nonumber \\
							&= 2e^{-1} \sinh (\varepsilon) \nonumber \\
							&\approx 0.7358 \cdot \varepsilon,
	\end{align}
	where $\varepsilon$ equals $10^{-50}$, $10^{-60}$, $10^{-100}$ or other values depending, respectively, on whether $d$ is the Hubble constant squared, the energy of the quantum vacuum, or any other relevant ratio under consideration. Since $\varepsilon \ll 1$, there is extreme fine tuning in all these cases.
	
\subsection{The maximum entropy principle applied to each of two random variables $G$ and $D$ in $\mathbb R^+$ that form a ratio.}
	
	As a second application of \eqref{Pmax}, assume that $X=G/D$, where $G>0$ and $D>0$ are independent random variables, with distributions chosen according to the MaxEnt principle. Assuming that the first moments of $G$ and $D$ are known, then $G\sim\mbox{Exp}(1/\mu_G)$ and $D\sim\mbox{Exp}(1/\mu_D)$, where 	$\mu_G=E(G)$ and $\mu_D=E(D)$ are two parameters that vary independently. Then $X$ has a ratio distribution with density
	\begin{align*}
			f_X(y;\mu) &= \int_0^\infty u f_D(u;\mu_D) f_G(uy;\mu_G)du\\
					&= \frac{1}{\mu} \cdot \frac{1}{(1+y/\mu)^2}
	\end{align*}
	for $y>0$ and $\mu=\mu_G/\mu_D$. This gives a tuning probability
	\begin{align}\label{A}
			P(X\in \ell_X) &= \frac{1}{\mu} \int_{x-\delta}^{x+\delta} \frac{1}{(1+y/\mu)^2}dy \nonumber\\
					&= 2\delta \cdot \frac{\mu}{(\mu+x-\delta)(\mu+x+\delta)}\\
					&= P(\varepsilon;\mu), \nonumber
	\end{align}
 	where $\varepsilon = \delta/x$. Since the tuning probability in (\ref{A}) only depends on $\mu_G$ and $\mu_D$ through their ratio $\mu = \mu_G/\mu_D$, according to \eqref{Pmax}, $P_{\mbox{\scriptsize max}}$ is obtained by maximizing (\ref{A}) with 		respect to $\mu>0$. This maximum is  
	\begin{equation}\label{B}
		\mu = \sqrt{(x-\delta)(x+\delta)} = x \sqrt{1-\varepsilon^2} \approx x.
	\end{equation}
	Inserting (\ref{B}) into (\ref{A}) we thus obtain the  upper bound of the tuning probability,
	\begin{align}\label{C}
			P_{\mbox{\scriptsize max}}(\varepsilon) &= P(x \sqrt{1-\varepsilon^2};\varepsilon) \nonumber\\
										&\approx 2x\varepsilon \cdot \frac{x}{4x^2}\\
										&= \frac{\varepsilon}{2} \nonumber.
	\end{align}
 	Note that (\ref{C}) is minuscule being slightly smaller than the corresponding upper bound in (\ref{ExMax1}), where the MaxEnt principle was applied directly to the distribution of the ratio $X$. Again, extreme fine tuning is seen whenever $\varepsilon$ is small.	
	
\subsection{The maximum entropy principle applied to a ratio $X=G/D$ in $\mathbb R$.}

	In the previous examples the only possibility considered was for gravity to be a nonnegative number. Were $G$ allowed to be negative, both repulsion and attraction would be possible. Were $G$ allowed to be also 0, gravity would be a neutral force. In this scenario, according to the first step, $G_\text{obs}$ must be thought of as a constant whose possible values constitute the whole real line $\mathbb R$. The same is true for the ratio $x=G_\text{obs}/d$ whether $d$ is negative or positive. In this example the MaxEnt principle will be applied to the distribution of $X$.
	
	In this section the MaxEnt principle is applied to $X$. Table \ref{maxent} says that, if the first two moments are known, the distribution that best explains $X$ is a normal $\mathcal N(\mu, \sigma^2)$. As in the previous example, $x$ can be regarded as the midpoint of $\ell_X$. This leads to a probability of the event $\{X \in \ell_X\}$ (step III):
	\begin{align}\label{Four}
		P(X\in \ell_X) &= \Phi[(x+\delta-\mu)/\sigma)] - \Phi[(x-\delta-\mu)/\sigma] = P(\mu,\sigma;\varepsilon),
	\end{align}
	where $\Phi(\cdot)$ is the standard normal distribution and $\varepsilon = \delta/x$. Generally, (\ref{Four}) is small for large $\sigma$ regardless of the value of $\mu$. However, by choosing $\mu=x$ in (\ref{Four}) and then letting $\sigma\to 0$, 	(\ref{Pmax}) attains the largest possible value
	\begin{equation}\label{Bound1}
		P_\text{max} = \sup_{\mu\in\mathbb R,\sigma>0} P(\mu,\sigma;\varepsilon) = 1.
	\end{equation}
	In this case, since $\mu/\sigma$ is unrestricted, the constant of proportionality $C$ in (\ref{PmaxEps}) is unbounded. This makes the tuning coarse.  
	
	In order to have a smaller value of $P_{\mbox{\scriptsize max}}$, additional restrictions to the class of prior distributions are necessary. For instance, under the assumption that neither positive nor negative values of $X$ should be favored a priori, enforcing 	$\mu=0$ is natural. Then (\ref{Four}) takes the form
	\begin{equation}\label{Seven}
		P(\sigma;\varepsilon) \approx \phi(x/\sigma) \frac{2 \delta}{\sigma},
	\end{equation}
	where $\phi(\cdot)=\Phi^\prime(\cdot)$ is the density function of a standard normal. This expression will have a very small upper bound, uniformly in $\sigma$, since
	\begin{align}\label{Bound2}
		P_{\max}(\varepsilon) &= \sup_{\sigma>0} P(\sigma;\varepsilon) \nonumber\\
						&\approx \frac{2\delta}{x} \max_{\sigma>0} \left\{ \phi(x/\sigma) \frac{x}{\sigma} \right\} \nonumber \\
						&= \frac{2\delta\phi(1)}{x} \\
						&\approx \frac{2\delta\cdot 0.242}{x} \nonumber\\
						&= 0.484\cdot \varepsilon, \nonumber
	\end{align}
	where the second equality is obtained by maximizing $\phi(x/\sigma)(x/\sigma)$ with respect to $\sigma>0$. This is equivalent to maximizing $\phi(z)z$ with respect to $z>0$, 	which achieves a maximum at $\phi(1) = e^{-1/2} / \sqrt{2\pi}  \approx 0.242$. 		See Figure {\ref{Gauss}} below, for an illustration of the bounds in (\ref{Bound1}) and (\ref{Bound2}). Again, we see extreme fine-tuning provided $0\notin \ell_G$ and $\varepsilon$ is small.
	
	\begin{figure}[h]
		\includegraphics[scale = 0.2]{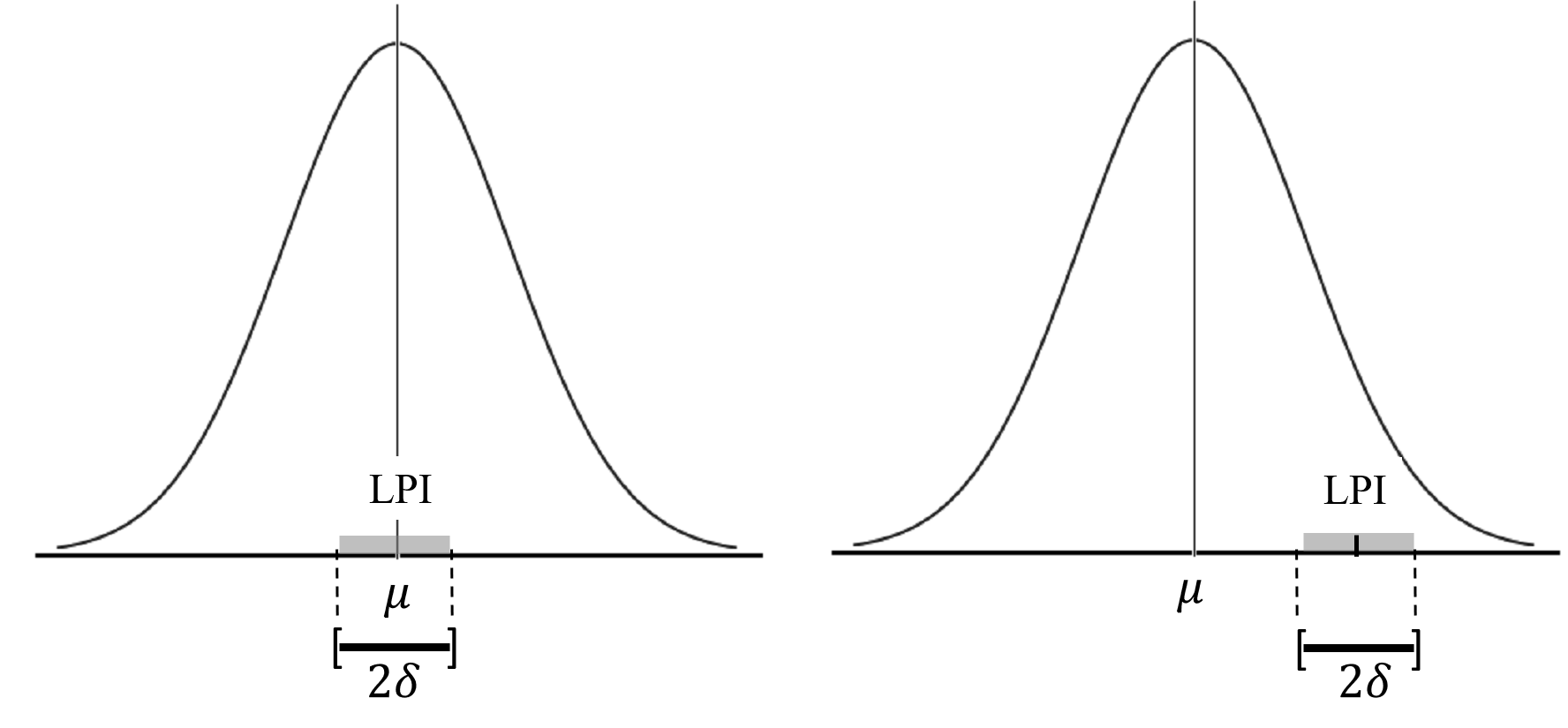}
		\centering
		\caption{When the variance $\sigma^2$ of the prior distribution of $X$  approaches 0, the normal distribution approaches a Dirac delta measure at $\mu$; thus $\mu \in \ell_X$, implies $P_{\mbox{\scriptsize max}} = 1$ (left). On the other hand, when 		$\mu \notin \ell_X$, where $\ell_X$ is the life-permitting interval (LPI), $P[\ell_G]$ will go to zero either when $\sigma \rightarrow 0$ or when $\sigma \rightarrow \infty$. Therefore $P_{\mbox{\scriptsize max}}$ is strictly less than 1 (right). For the figure at 		the left, the tuning is coarse, whereas for the figure at the right it is fine.}\label{Gauss}
	\end{figure}
	
\subsection{The maximum entropy principle applied to each of two random variables $G$ and $D$ in $\mathbb R$ that form a ratio.}	

	We now turn to the second approach of choosing the distribution $F$ of $X=G/D$, by using the MaxEnt principle to $G$ and $D$ separately, where $G$ and $D$ are both allowed to take on negative and positive values. In order not to favor either positive or 	negative values of $G$ or $D$, assume that $G\sim \mathcal N(0,\sigma_G^2)$ and $D\sim \mathcal N(0,\sigma_D^2)$. Then $X$ has a Cauchy distribution with density
	\begin{align*}
			f_X(y;\sigma) &= \int_{-\infty}^\infty |z|f_D(z;\sigma_D)f_G(yz;\sigma_G)dz\\
						&= \frac{1}{\sigma\pi} \frac{1}{(1+(y/\sigma)^2)},
	\end{align*}
	for $y\in\mathbb R$, with $\sigma=\sigma_G/\sigma_D$. This gives a tuning probability
	\begin{align}\label{D}
			P(\sigma;\varepsilon) &= \int_{x-\delta}^{x+\delta} f_X(z;\sigma)dz \nonumber \\
							&\approx  2\delta f_X(x;\sigma)\\
							&=  \frac{2x\varepsilon}{\sigma} \frac{1}{\pi(1+(x/\sigma)^2)} \nonumber
	\end{align}
	that only depends on $\sigma_G$ and $\sigma_D$ through their ratio $\sigma$. Maximizing (\ref{D}) with respect to $\sigma>0$ is equivalent to taking the maximum with respect to $z=x/\sigma>0$. Consequently
	\begin{align*}
			P_\text{max}(\varepsilon) &\approx  \frac{2\varepsilon}{\pi} \max_{z>0} \frac{z}{1+z^2}\\
								&= \frac{\varepsilon}{\pi}.
	\end{align*}
	Interestingly, the Cauchy distribution itself (with scale parameter $\sigma$) is MaxEnt over $\mathbb R$ under the restriction that $E(\ln (1 +X^2/\sigma^2)) = 2 \ln 2$ \cite{ParkBera2009}.
	
\subsection{The maximum entropy principle applied to each of two constants $G$ in $\mathbb R$ and $D$ in $\mathbb R^+$, that form a ratio.}
	
	Another possibility is when $G\in \mathbb R$ and $D>0$ are chosen to have MaxEnt distributions $G\sim \mathcal N(\mu_G,\sigma_G^2)$ and $D\sim \mbox{Exp}(1/\mu_D)$. As before, we assume $\mu_G = 0$.  Then $X=G/D$ will have a symmetric ratio 	distribution with scale parameter $\sigma = \sigma_G/\mu_D$. The tuning probability is obtained similarly as in (\ref{D}), although the distribution of $X$ is no longer Cauchy.   	
	
\subsection{The maximum entropy principle with an upper bound on prior information}

Recall from Section 4.3 that a normal MaxEnt prior distribution $N(\mu,\sigma^2)$ was assumed for a real-valued ratio $X=G/D$ of the constant of gravity $G$ and another constant of nature $D$. The associated tuning probability $P_\text{max}$ differs a lot depending on whether $\mu=0$ is enforced or not. The signal-to-noise ratio $\text{SNR}=\mu^2/\sigma^2$ quantifies the amount of information that the prior distribution carries about $X$, on a relative scale. With no restriction on $\mu$, the prior distribution could have an arbitrarily high  $\text{SNR}$, resulting in $P_\text{max}=1$, whereas $\mu=0$ corresponds to a requirement $\text{SNR}=0$. In between these two extremes, we will assume that the signal-to-noise ratio has a finite upper bound
	\begin{align}
		\text{SNR} \le \text{SNR}_\text{max} \ll \varepsilon^{-2},
	\end{align}
where $\varepsilon$, the half length of the life-permitting interval $\ell_X$ divided by the modulus of its mid point, is a small number. Then it can be shown that the degree of tuning for a real-valued constant of nature, or a real-valued ratio of two such constants, is given by  
\begin{equation}\label{XXX}
	P_\text{max} = \max \left\{P(\mu,\sigma;\varepsilon); \,\, \frac{\mu^2}{\sigma^2} \le \text{SNR}_\text{max} \right\} \approx 2\varepsilon \sqrt{\text{SNR}_\text{max}} \phi(0) \ll 1,
\end{equation}
where $P(\mu,\sigma;\varepsilon)$ is the tuning probability defined in \eqref{Four}, with $\delta=x\varepsilon$ and $x$ the observed value of $X$.

\section{Example 2: Amplitude of primordial fluctuations}

	The amplitude of primordial fluctuations,  $q = Q_\text{obs} \approx 2\times 10^{-5}$, is here the dimensionless value of interest \cite{Barnes2011}.  Notice that in contrast to Example 1, $q$ is here a constant of nature rather than a ratio between two constants of nature. Rees writes \cite[p.~128]{Rees2000}:
	\begin{quotation}
		``If [$Q_\text{obs}$] were smaller than $10^{-6}$, gas would never condense into gravitationally bound structures at all, and such a universe would remain forever dark and featureless, even if its initial `mix' of atoms, dark energy and radiation were 			the same as our own. 
		
		On the other hand, a universe where [$Q_\text{obs}$] were substantially larger than $10^{-5}$ -- where the initial `ripples' were replaced by large-amplitude waves -- would be a turbulent and violent place. Regions far bigger than galaxies would 			condense early in its history. They wouldn't fragment into stars but would instead collapse into vast black holes, each much heavier than an entire cluster of galaxies in our universe... Stars would be packed too close together and buffeted too 				frequently to retain stable planetary systems.''
	\end{quotation}

	 Adams gives an even larger life-permitting interval than Rees: $10^{-6} \leq Q_\text{obs} \leq 10^{-2}$ \cite{Adams2019}. But as we will see, there is little difference in our overall conclusion that the amplitude of primordial fluctuations is coarsely tuned. Following the three-step procedure for calculating $P_\text{max}$, the first step is to determine the right sample space. 
	
	The amplitude of the primordial fluctuations could be any nonnegative number; in other words, there seems to be no mathematical restriction to impose that the sample space of $Q$ must be bounded above. On the other hand, were $Q$ large, the FRW assumption (e.g., homogeneity and isotropy) would  be violated, which implies that the universe obtained would be so different to ours that life as we know it would not be possible \cite[see, e.g.,][pp.~118, 128]{Rees2000}. Therefore such universe would not be tuned for life. As such, large values of $Q$ impose a boundary on the life-permitting interval, not on the sample space of $Q$. Since we already have an upper bound of the life-permitting interval, the FRW assumption imposes no further restriction on this interval. The conclusion is that the sample space of $Q$ is $\mathbb R^+$. 
	
	Thus, from Table \ref{maxent}, we see that $Q \sim \text{Exp}	(1/\mu)$, provided the mean $\mu$ is given. The third step is to calculate the probability of the life-permitting interval, so let's take $\ell_Q = (10^{-6}, 10^{-5})$ 
	\begin{align}\label{ProbQ}
		 P(Q \in \ell_Q) = \exp \left(-10^{-6}/\mu \right) - \exp \left(-10^{-5}/\mu \right) = P(\mu),
	\end{align}
	that depends on $\mu$. In this case, according to \eqref{Pmax}, the maximal value is $P_\text{max} = \max_{\mu >0} P(\mu)$. This maximization is similar to the one given in Section 4.1. Putting $a=10^{-6}$ and $b=10^{-5}$, the maximum tuning 		probability is obtained for
	\begin{align}\label{muQ}
		\mu = \frac{b-a}{\log(b/a)} = \frac{9\times 10^{-6}}{\log(10)}.
	\end{align}
	Inserting (\ref{muQ}) into (\ref{ProbQ}) we obtain a rather large probability.
	\begin{align}
		P_{\mbox{\scriptsize max}} &= 10^{-1/9} - 10^{-10/9}  \nonumber\\
						&\approx 0.697.
	\end{align}
	This is surprising given that there are claims of a much higher degree of tuning \cite{TegmarkRees1998}. In fact, our choice of $\ell_Q$ is extremely generous in favor of fine-tuning, taking into account that $Q_\text{obs} \approx 2\times 10^{-5} \notin (10^{-6}, 10^{-5}) = \ell_Q$! This reveals that there looks to be a coarse tuning on the amplitude of the primordial fluctuations according to the MaxEnt model. From a mathematical viewpoint, the coarse tuning of $Q$ is not too surprising though, since the upper limit of $\ell_Q$ is one order of magnitude larger than its lower limit.

\section{Discussion}

\subsection{Summary}

In this paper a Bayesian statistical procedure has been devised for calculating an upper bound $P_{\mbox{\scriptsize max}}$ for the probability that a constant of nature, or the ratio of two constants of nature, belongs to a life permitting interval. This upper bound is invariant with respect to change of physical units, and under certain assumptions it is small whenever the life-permitting interval is small on a relative scale, corresponding to a small value of $\varepsilon$. We obtain $P_{\mbox{\scriptsize max}}$ through a three-steps procedure, where I) the sample space is determined, II) the finite-dimensional class of distributions, of the constant of nature or the ratio between constants of nature, is found for this sample space by applying the maximum entropy principle, and III) the tuning probability is maximized over this finite-dimensional class of distributions. The overall conclusion is that the probability is proportional to the relative width of the life-permitting interval, which will be small for intervals of narrow width, unless the prior distribution carries a lot of information and is concentrated within this region.

\subsection{Choice of sample space}

Recall that the first step of our tuning approach is to choose a sample space $\Omega$ for a constant of nature or a ratio of such constants. This raises the question of what should influence the choice of $\Omega$. For instance, for the gravitational constant $G$ of Section 4, and the choice the whole real line as the set of possible values of $G$, as in Sections 4.3 and 4.4, some concern may appear when $G<0$ taking into account that in this case flat space is explosively unstable. However this does not rule out the possibility of $G<0$, it only rules out the possibility of a universe of the kind we know. In particular for the tuning problem, it would exclude the possibility of carbon-based life. But this does not entail that a negative gravitational constant cannot be possible. To see this more clearly, Cline, Jeon, and Moore argue that {\sl under our current theories} certain constraints must be imposed on the physical parameters  \cite{ClineJeonMoore2004}. This is true when we are evaluating our theories or applying them to study the natural world {\sl within}, not the natural world {\sl as such}.  For the latter, it is not theories about the constants that we are studying, but the constants themselves. Therefore, in the most strict sense, such theories do not {\sl determine} the support of the random variables nor their realization in the constants of nature as we know them.  Cosmological tuning ---why the physical constants have the values they have--- is about the world as such, not about the world within. Were theories in focus, fine-tuning would be studied for theories, not for natural constants. 

Physical parameters can be measured and quantified, whereas theories are abstractions. For this reason, by inductive reasoning, it is physical parameters that determine which theories are acceptable. It is also important to have in mind that natural constants are actually constants, not observations of random variables.  Nevertheless, since all the Bayesian fine tuning enterprise dwells on the assumption of random variables whose observations are the natural constants we observe, it might be artificial to restrain the values that such random variables can take for reasons other than mathematical. This is to say that if natural constants are more fundamental than physical theories, it seems reasonable that a priori constraints on such constants are dictated by mathematics rather than theories.

Finally, it is not only physical theories but also observations that possibly could restrict a priori assumptions on the constants of nature. The latter is in fact an instantiation of the weak anthropic principle.  As we argued in Section 3, attempts to estimate hyperparameters based on observations are not the right approaches to measure the probabilities of tuning, since they are adding bias to the measurements of probabilities of life-permitting intervals.

\subsection{Choice of prior and MaxEnt approach}

It is important to notice that our approach goes beyond total ignorance and can incorporate whatever knowledge is at hand, provided a MaxEnt distribution exists under the relevant restrictions. In practice this means that if we are using Lagrange optimization to obtain the MaxEnt distribution, such knowledge will take the form of constraints --- additional terms with their respective Lagrange multipliers. Thus the addition of partial knowledge can be incorporated into our MaxEnt approach.

In fact, for the sample spaces as considered in the examples, other distributions are possible under different restrictions. For instance, Park and Bera list eight MaxEnt distributions for $(-\infty, \infty)$, and six MaxEnt distributions for $[0,\infty)$, corresponding to different restrictions \cite{ParkBera2009}. Of course, there are many other ways of choosing restrictions on the prior distributions. This highlights the fact that the MaxEnt principle does not imply that the distribution of the parameter needs to be totally unknown. Rather, it means that the distribution is maximally unbiased given whatever knowledge there is at hand. Indeed, if the distribution is known, it must correspond to the MaxEnt distribution, conditioned on whatever is making it known.

In order to describe in more detail how additional knowledge of constants of nature influences the MaxEnt approach, consider equation  \eqref{Pmax}: Additional knowledge may on one hand increase the dimensionality $d$ of the parameter vector $\theta=(\theta_1,\ldots,\theta_d)$, if, for instance, each component of $\theta$ corresponds to a different aspect of the distribution of the constant of nature $X$ (or the distribution of the ratio $X$ of two constants of nature), such as the mean, variance, third moment, truncation limits etc. Secondly, the more information that is available about the distribution of $X$, the more the $d$-dimensional parameter space $\Theta$ is shrunk. In particular, total knowledge of the distribution of $X$ implies that $\Theta$ has only one element, say $\theta^\ast$, in which case $P^*=F(\ell_X;\theta^*)$ is obtained from the only permissible distribution of $X$; the one that corresponds to the known hyperparameter $\theta^*$.     

Now, if the "known" fine tuning probability $P^*$ does not correspond to $P_\text{max}$, that was obtained in \eqref{Pmax} with less knowledge about the hyperparameter $\theta$, the bias produced by $P^*$ must be accounted for.  A very efficient way to keep accountability is by calculating the active information $I^+ = \log(P^*/P_\text{max})$
 \cite{DembskiMarks2009b, DiazMarks2020a, DiazSaenzRao2020}. $I^+$ measures the amount of information added or subtracted when $P^*$ is used instead of $P_\text{max}$. If $I^+$ is positive, $P^*$ is overfitting for $\ell_X$ with respect to $P_\text{max}$; if $I^+$ is negative, $P^*$ is underfitting the life-permitting interval with respect to $P_\text{max}$; and if $I^+=0$, $P^*$ the extra assumptions used to calculate $P^\ast$ do not contribute anything in terms of cosmological tuning.

\subsection{Outlook}

The purpose of the examples was to illustrate as simply as posible the mathematical approach to find expressions for the upper bound $P_{\mbox{\scriptsize max}}$ of the tuning probability. However, these upper bounds are accurate, given the assumed class $\mathcal F$ of prior distributions and the observed life permitting interval $\ell_X$. In particular we think that, if the constants are really fundamental, they cannot be determined by theories, and it is theories that have to accommodate to fundamental constants. Only other constant could constrain the existing ones (however, we do not have a theory of everything that allows us to do so). For these reasons we regard our approach as promising for future research.

On the other hand, a more epistemological approach can limit the sample spaces of constants of nature, or ratios of constants of nature, according to restrictions set by theories. Therefore, this paper sets the stage for a fruitful research project in several aspects:

\begin{enumerate}
	\item To calculate the level of tuning for other parameters, constants of nature, and boundary conditions. For instance, from Example 1, the ratio $x=\Lambda_\text{vac}/\Lambda_\text{bare}$ of the two terms involved in the cosmological constant is contained within an interval $\ell_X$ that is centered around -1, with a relative half length of $\varepsilon = 10^{-120}$.  
	\item To study tuning probabilities under different assumptions that may be imposed by theories.
	\item To find upper bounds for the joint tuning probability of several constants of nature. This would in turn provide a lower bound for the expected number of parallel universes needed in order to obtain one that permits life.
	\item To estimate the tuning probability by means of the signal-to-noise ratio, in order to obtain an upper bound on the amount of information that the prior distribution carries about a constant of nature, or a ratio of such constants. In Section 4.6 this was done in a context where the first two moments of the prior distribution are specified. This approach can be generalized, using other measures of information of the prior distribution.   
\end{enumerate}

Finally, as mentioned in Section 1, the actual lengths of the life-permitting intervals might change in the future, and thereby affect the values  of $P_\text{max}$. This is not a weakness of our method, since it will serve to quantify the amount of tuning for a specific interval. The very nature of the scientific enterprise will also produce new constants of nature, with associated life-permitting intervals. In fact, as we saw in the examples, different theories do produce different intervals. When new intervals or new parameters arrive, our method will give an accurate upper bound to the probability of tuning, conditionally on this new and updated knowledge.

\section{Acknowledgements}

The authors are thankful to Luke Barnes for his suggestions, as well as to Aron Wall for his observations on the first version of this paper. We are also grateful to an anonymous reviewer for excellent comments which greatly improved this paper.

\bibliographystyle{ieeetr}

\end{document}